\def\a{\alpha}\def\d{\delta}\def\e{\epsilon}
\def\f{\phi}\def\g{\gamma}\def\h{\theta}
\def\m{\mu}\def\n{\nu}\def
\p{\pi}\def\r{\rho}\def\s{\sigma}

\def\D{\Delta}
\def\Om{\Omega}

\def\de{\partial}
\def\ha{{1\over 2}}

\def\({\left(}\def\){\right)}\def\[{\left[}\def\]{\right]}

\def\mn{{\mu\nu}}\def\ij{{ij}}\def\ba{\bar\a}

\def\bc{boundary conditions }\def\ssy{spherically symmetric }

\def\af{asymptotically flat }
\def\fe{field equations }\def\bh{black hole }\def\coo{coordinates }

\def\sch{Schwarzschild }
\def\RN{Reissner-Nordstr\"om }

\def\GB{Gauss-Bonnet }\def\ab{asymptotic behavior }
\def\wrt{with respect to }

\def\section#1{\bigskip\noindent{\bf#1}\smallskip}

\def\nota{\footnote{$^\dagger$}}

\font\smalll = cmr7

\def\PL#1{Phys.\ Lett.\ {\bf#1}}\def\CMP#1{Commun.\ Math.\ Phys.\ {\bf#1}}

\def\PR#1{Phys.\ Rev.\ {\bf#1}}\def\CQG#1{Class.\ Quantum Grav.\ {\bf#1}}
\def\NP#1{Nucl.\ Phys.\ {\bf#1}}
\def\JMP#1{J.\ Math.\ Phys.\ {\bf#1}}
\def\PRS#1{Proc.\ R. Soc.\ Lond.\ {\bf#1}}
\def\JoP#1{J.\ Phys.\ {\bf#1}}

\def\AoP#1{Ann.\ Phys.\ {\bf#1}}

\def\ref#1{\medskip\everypar={\hangindent 2\parindent}#1}
\def\beginref{\begingroup
\bigskip
\centerline{\bf References}
\nobreak\noindent}

\newfam\amsfam
\font\tenams=msam10 \textfont\amsfam=\tenams
\mathchardef\ls="382E\mathchardef\gs="3826
\input epsf
\baselineskip=12pt

\def\br{\bar r}
{\nopagenumbers
\line{}
\vskip20pt
\centerline{\bf Dyonic black holes in nonlinear electrodynamics}
\centerline{\bf from Kaluza-Klein theory with a Gauss-Bonnet term}

\vskip60pt

\centerline{{\bf S. Mignemi}\nota{e-mail:\ smignemi@unica.it}}
\vskip5pt
\centerline {Dipartimento di Matematica, Universit\`a di Cagliari}
\centerline{via Ospedale 72, 09124 Cagliari, Italy}
\smallskip
\centerline{and}
\centerline{INFN, Sezione di Cagliari, Cittadella Universitaria,}
\centerline{09042 Monserrato, Italy}

\vskip80pt
{\noindent\centerline{\bf Abstract}}
\vskip5pt
Five-dimensional Kaluza-Klein theory with an Einstein-Gauss-Bonnet Lagrangian induces nonlinear corrections
to the four-dimensional Maxwell equations, which however remain second order.
Although these corrections do not have effect on the purely electric or magnetic monopole solutions for pointlike
charges, they affect the dyonic solutions, smoothing the  electric field at the origin for positive values
of the \GB coupling constant.
We investigate these solutions in flat space, and then extend them in the presence of a minimal coupling to gravity,
obtaining exact charged black hole solutions that generalize the \RN metric.

\vfil\eject}

\section{1. Introduction}
It is well known that Maxwell equations can be generalized in a non-linear way, adding to the Lagrangian
higher powers of the invariants constructed from the electromagnetic field. Well-known examples are the corrections due to
quantum electrodynamics that were proposed by Heisenberg and Euler [1] or the highly non-linear Born-Infeld Lagrangian
[2] and their generalizations by Pleba\'nski [3]. These generalizations still yield second order field equations,
but can give rise to solutions with regular electric or magnetic field [3].

Nonminimal coupling of the Maxwell equations to the gravitational field is instead more difficult, if one requires that the
field equations remain second order and linear in the second derivatives of the electromagnetic potential and of the metric tensor.
This problem has been studied in general in [4]. It is notable that a simple example of a model obeying this property can be
obtained by dimensional reduction of a Kaluza-Klein (KK) theory containing a Gauss-Bonnet (GB) contribution [5-8].

We recall that KK theories [9,10] provide a unification of
general relativity with electromagnetism based on the assumption that spacetime is five-dimensional
and the fifth dimension is not observable because it is curled in an extremely small circle. However,
in higher dimensions the Einstein-Hilbert Lagrangian is not unique, and one may add to it a GB term,
which would not be effective in four dimension, since in that case it reduces to a total derivative.
GB terms were shown in [11] to give the most general corrections to higher-dimensional gravity leading to
second order field equations and compatible with some natural assumptions.

The introduction of this term in the five-dimensional Lagrangian permits to obtain by dimensional reduction to four dimensions
a model whose predictions differ from those of the Einstein-Maxwell (EM) theory, giving rise to the possibility of an
indirect evidence of the existence of a fifth dimension.
In fact, the dimensionally reduced theory contains corrections to the EM coupling that are of the kind discussed in [4].
Moreover, they provide nonlinear modifications of the pure electromagnetic lagrangian, that give rise to
corrections of the standard electrodynamics [6].
Although these corrections can be considered as a special case of Pleba\'nski's nonlinear electrodynamics [3], and in particular
of its simplified version proposed in [12], their properties are rather peculiar, due to the particular combination of coefficients
in the Lagrangian. For example, the purely electric or magnetic solutions of the Maxwell equations are not modified.
It follows that, although regular solutions can be obtained for more general quadratic electrodynamics coupled to gravity [13-15],
this is not the case for this model.

These facts are particularly relevant in relation with uniqueness and no-hair theorems for black holes.
These theorems state that the only \ssy \af solution of the EM theory is the RN metric [16].
However, if nonlinear electromagnetic terms, like those of Born-Infeld [17] or Pleba\'nski [13-14], or extra fields with nonminimal coupling,
like the dilaton [18-19], are added to the standard EM action, the theory will exhibit different solutions.
Also the generalization to Yang-Mills fields can give rise to nontrivial solutions [20].

In fact, the solutions of the five-dimensional Einstein-GB theory have been studied from a higher-dimensional point of view in ref.~[8],
where it was shown that the effect of the GB term is only detectable through the coupling of electrodynamics to the gravitational field.
However, the case of dyonic solutions was disregarded in that paper. As we shall see,
dyonic solutions of the standard Maxwell equation are modified, even in the absence of the gravitational field,
due to the nonlinear terms present in the field equations. Dyons were introduced in ref.~[21] and have found many
application in grand unified theories. Especially interesting are also their implications on the properties
of charged black holes, in particular in relation with uniqueness and no-hair theorems.
The coupling of our dyonic solution with gravity of course modifies the standard RN black holes, giving
another example of the failure of the uniqueness theorems in case of nontrivial couplings.

In this paper, we describe exact flat-space dyonic solutions of the nonlinear Maxwell equations derived from GB-KK theory
and show that solutions with everywhere regular electric field are possible.
We then investigate the effect of these configurations on the \bh solutions of general relativity if the electromagnetic
field is minimally coupled, obtaining an exact solution. We briefly discuss its properties and thermodynamical
parameters.

However, we shall not consider the nonminimal couplings with the gravitational field arising from the dimensional
reduction of the GB Lagrangian, since this problem is more involved and will therefore be studied separately [22].
\goodbreak

 \section{2. The dyonic solution in flat space}
We consider a five-dimensional Einstein-Gauss-Bonnet theory, with action
$$I=\int\sqrt{-g}\ d^5x(R+\a S),\eqno(1)$$
where $\a$ is a coupling constant of dimension (lenght)$^2$, $R$ is the Ricci scalar and $S$ the Gauss-Bonnet term,
$S=R^{\m\n\r\s}R_{\m\n\r\s}-4R^{\m\n}R_{\m\n}+R^2$.

We use the simple ansatz\footnote{$^\dagger$}{Greek indices run from 0 to 4, Latin indices from 0 to 3.} [5]
$$g_\mn=\(\matrix{g_\ij+g^2A_iA_j&gA_i\cr gA_j&1}\),\eqno(2)$$
where $A_i$ is the Maxwell potential and $g$ a coupling constant.
Discarding total derivatives, the action (1) reduces to [1-3]
$$I=\int\sqrt{-g}\ d^4x\[R-{g^2\over4} F^\ij F_\ij+{3\a g^4\over16}\Big[(F^\ij F_\ij)^2-2F^{ij}F_{jk}F^{kl}F_{li}\Big]- {\a g^2\over2}L_{int}\],\eqno(3)$$
where
$$L_{int}=F^\ij F^{kl}(R_{ijkl}-4R_{ik}\d_{jl}+R\,\d_{ik}\d_{jl}),\eqno(4)$$
and $F_\ij=\de_iA_j-\de_jA_i$.
This model of electromagnetism modified with nonlinear terms has been previously considered in ref.~[4]. The
Einstein-Maxwell coupling (4) has also been investigated in ref.~[8].

In this section we shall consider only the electromagnetic field in flat spacetime, neglecting gravity, since we
are mainly interested in the nonlinear modifications of the Maxwell theory.
The solutions of the EM theory (3) will be discussed in the following section.

The electromagnetic sector of the action (3) then reduces to [2]
$$I_{em}=\int d^4x\ \(-{g^2\over4}F^\ij F_\ij+{3\a g^4\over16}\Big[(F^\ij F_\ij)^2-2F^{ij}F_{jk}F^{kl}F_{li}\Big]\).\eqno(5)$$
The \fe derived from (4) read
$$\(1-{3\a g^2\over2}F^2\)\de_jF_{ji}+3\a g^2\,\de_j(F_{jk}F_{kl}F_{li})=0.\eqno(6)$$
and contain derivatives of the potential $A_i$ not higher than second order.
Of course, the field $F_\ij$ also satisfies the Bianchi identities, $\de_{(i}F_{jk)}=0$.

It is easy to see that for purely electric or magnetic solutions the terms coming from the GB correction
give no contribution [2,4], in contrast with most nonlinear models of electrodynamics [3].
However, let us consider a \ssy dyonic solution, whose potential is given in spherical \coo by
$$A=a(r)\,dt+ P\cos\h\,d\f.\eqno(7)$$
In an orthogonal frame one has
$$F_{01}=a'(r),\quad F_{23}={P\over r^2},\eqno(8)$$
where $'=d/dr$. Clearly, $F_{23}$ satisfies (6).

To find the solution for the electric potential $a(r)$, it is convenient to write the action in terms of it and perform the variation.
After integration on the angular variables, the action is proportional to
$$I_{em}=\int r^2dr\[a'^2-{P^2\over r^4}+3\a g^2{P^2a'^2\over r^4}\],\eqno(9)$$
and its variation gives
$$\[r^2\(1+3\a g^2{P^2\over r^4}\)a'\]'=0.\eqno(10)$$
Therefore,
$$a'=F_{01}={Q\over r^2\(1+3\a g^2{P^2\over r^4}\)}={r^4\over r^4+3\a g^2P^2}\ {Q\over r^2},\eqno(11)$$
with $Q$ an integration constant, that can be identified with the electric charge.
The potential can be obtained by integration.

It follows that in this model the electric field of a point charge is distorted in the presence of a magnetic monopole; in
particular, for $\a<0$ it diverges at $r=(3|\a|g^2P^2)^{1/4}$, while for $\a>0$ it is regular everywhere.
However, the magnetic field (7) is still singular at the origin.
In the limits $P=0$ or $Q=0$ one recovers the standard solutions.

\section{3. The coupling to gravity}
We now introduce gravity, to see the effects of nonlinear electromagnetism on \ssy \bh solutions.
However, we neglect the nonminimal EM coupling (4), since in this way we can obtain exact solutions.
The inclusion of this term will be investigated elsewhere [22].
Moreover, contrary to ref.~[4], we consider the solutions of the four-dimensional effective theory, rather than those of
the five-dimensional one, since they admit a more transparent interpretation.
In the following, in order to obtain the standard normalization, we shall set $g^2=4$ and
define $\bar\a={\a/4}$, so that $\a g^2=\ba$.

\goodbreak

We seek for spherically symmetric solutions of the form
$$ds^2=-e^{2\n}dt^2+e^{2\m}dr^2+e^{2\r}d\Om^2,\eqno(12)$$
$$A=a(r)\,dt+ P\cos\h\,d\f.\eqno(13)$$
Like in the flat case, we calculate the \fe by substituting this ansatz into the action and performing the variation. We obtain
$$I=2\int dr\[(2\n'\r'+\r'^2)\e^{\n-\m+2\r}+e^{\n+\m}+a'^2e^{-\n-\m+2\r}-P^2e^{\n+\m-2\r}-3\ba P^2a'^2e^{-\m-\n-2\r}\].\eqno(14)$$
In the gauge $e^\r=r$ the \fe stemming from (14) read
$$2{\n'\over r}+{1\over r^2}-{e^{2\m}\over r^2}+a'^2e^{-2\n}+{P^2\over r^4}e^{2\m}+3\ba a'^2{P^2\over r^4} e^{-2\n}=0,\eqno(15)$$
$$-2{\m'\over r}+{1\over r^2}-{e^{2\m}\over r^2}+a'^2e^{-2\n}+{P^2\over r^4}e^{2\m}+3\ba\,a'^2{P^2\over r^4} e^{-2\n}=0,\eqno(16)$$
$$\[r^2e^{-\n-\m}\(1+3\ba\,{P^2\over r^4}\)a'\]'=0.\eqno(17)$$
Combining (15) and (16), we get
$$\n'+\m'=0\eqno(18).$$
Hence, for \af solutions, $\m=-\n$ and, integrating (17),
$$a'={Qr^2\over r^4+3\ba P^2},\eqno(19)$$
with $Q$ an integration constant.
Substituting in (15), one can rearrange as
$$(re^{2\n})'=1-{P^2\over r^2}-{Q^2r^2\over r^4+3\ba P^2}\approx1-{P^2+Q^2\over r^2}-{3\ba P^2Q^2\over r^6}+\dots\eqno(20)$$
which displays order-$\ba$ corrections to the corresponding equation for the RN metric function.
\goodbreak
\bigskip
\centerline{\epsfysize=4truecm\epsfbox{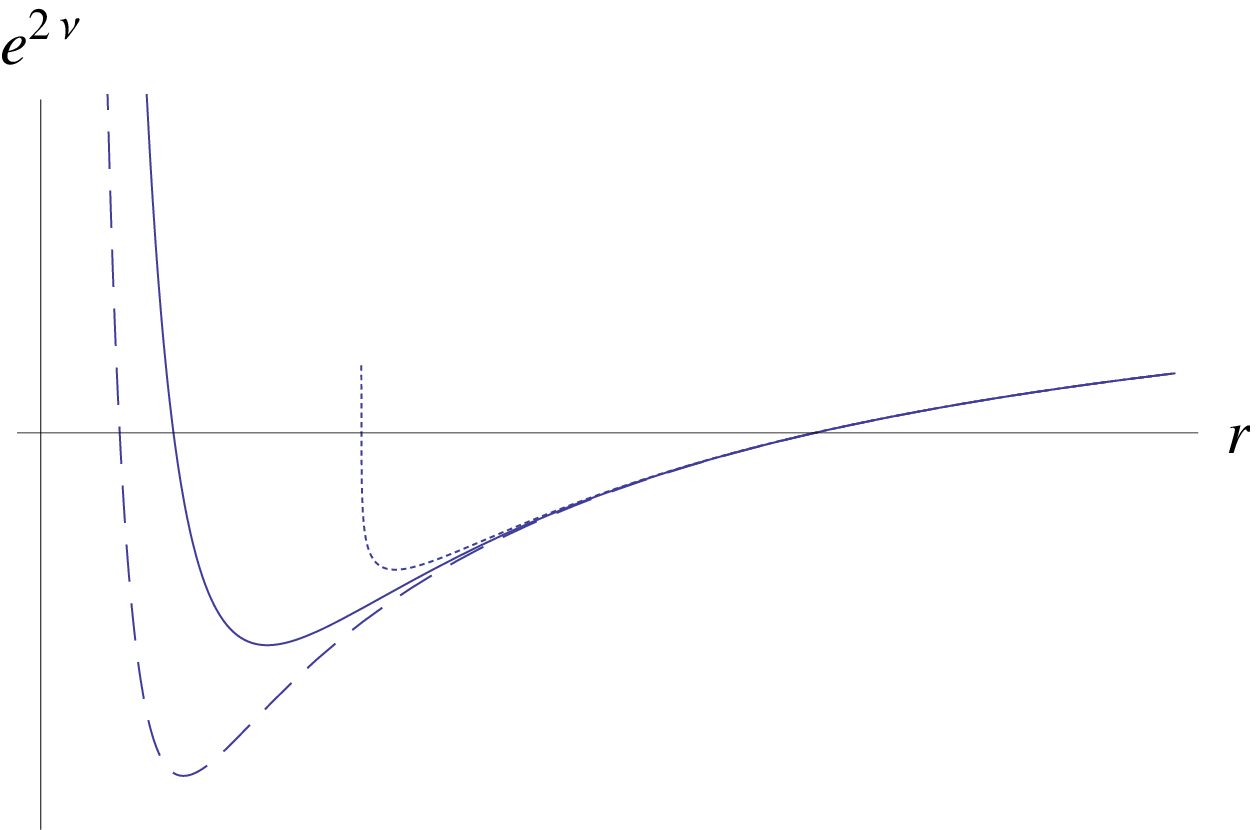}\qquad\epsfysize=4truecm\epsfbox{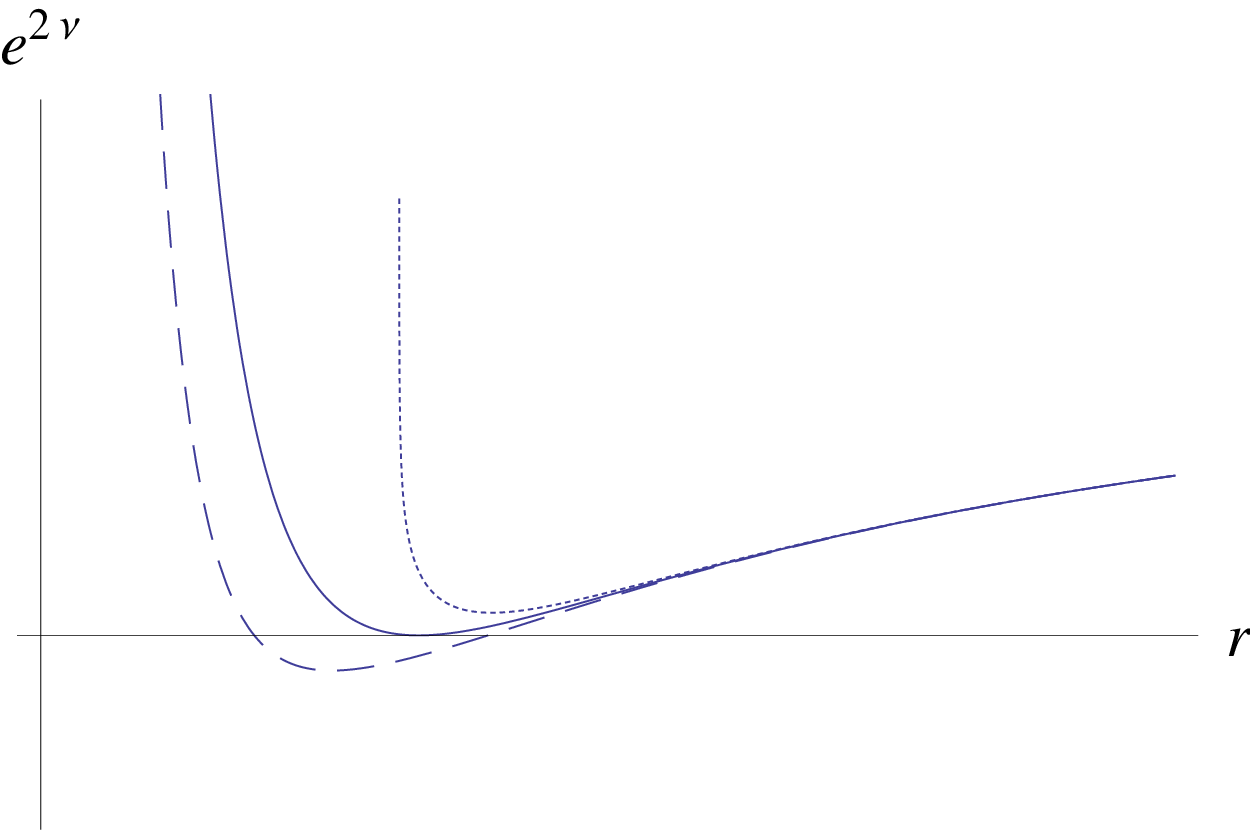}}
\medskip
\baselineskip10pt{\noindent{\smalll Fig.\ 1: The metric function $\scriptstyle{e^{2\n}}$ for generic (left panel) and near-extremal black holes (right panel).
In the right panel, we have chosen values of the parameters that are extremal for the RN black hole.
The continuous lines show the RN solution, the dashed line the $\scriptstyle{\a>0}$ solution (21) and the dotted line the
$\scriptstyle{\a<0}$ solution (31).
The singularity at $\scriptstyle{r=r_0}$ occurring when $\scriptstyle{\a<0}$ is clearly visible.}}\par
\bigskip
\baselineskip12pt
 Eq.~(20) can be solved exactly.
 Let us first consider the case $\a>0$. Setting $\g=\sqrt{3\ba P^2}$ and choosing suitable boundary conditions,
 the solution is
$$e^{2\n}=\ 1-{2M\over r}+{P^2\over r^2}+{Q^2\over2\sqrt{2\g}\,r}\[\p+\arctan\(1-{\sqrt2\,r\over\sqrt\g}\)-\arctan\(1+{\sqrt2\,r\over\sqrt\g}\)
+\ha\log{r^2-\sqrt{2\g}\,r+\g\over r^2+\sqrt{2\g}\,r+\g}\].\eqno(21)$$
This metric exhibits some similarity with the so-called geon solution of the Born-Infeld nonlinear electromagnetism coupled to gravity [11].\footnote{$^\ddagger$}
{Curiously, a similar solution has also been obtained in a rather different EM model with nonlinear terms [23].}
For small $\g$, it gives a slight correction to the RN solution, which however is relevant for the uniqueness theorems.
In Fig.~1 some solutions are depicted together with the corresponding RN metric function.

The \ab of (21) reproduces that of the RN metric,
$$e^{2\n}=1-{2M\over r}+{P^2+Q^2\over r^2}+o\({1\over r^3}\),\eqno(22)$$
and one can identify $M$ with the mass, $Q$ and $P$ with the electric and magnetic charge, respectively.
Instead for $r\to0$ the behavior is different from that of RN,
$$e^{2\n}\sim{P^2\over r^2}-\(2M-{\p Q^2\over2\sqrt{2\g}}\){1\over r}+o\({1}\).\eqno(23)$$

In particular, the ${1\over r}$ term becomes repulsive near the origin for $M<{\p Q^2\over4\sqrt{2\g}}$.
The departure from the RN behavior are therefore greater for small $r$. The term proportional to ${1\over\sqrt\g}$
arises because we have fixed the \bc so that $M$ is the mass of the solution. It can be useful to define an effective mass near
the singularity as $m=M-{\p Q^2\over4\sqrt{2\g}}$.

The curvature scalar is given by $R={4\g^2Q^2\over(r^4+\g^2)^2}$ and is regular everywhere, but, in contrast with the RN solution
does not vanish.
However also in our case a curvature singularity occurs at the origin, since
$$R_{ijkl}R^{ijkl}\sim{56P^4\over r^8}-{96mP^2\over r^7}+{48m^2\over r^6}+o\({1\over r^5}\).\eqno(24)$$
The leading order term in (24) depends only on $P$ and not on $Q$.

The causal structure depends on the values of the parameters $M$, $P$ and $Q$ that characterize the solution.
Due to the nontrivial form of the metric, a general discussion can be made only numerically. However, if $\g\ll1$, as
arguable on physical grounds, the solution should not differ much from the RN metric and one can resort to a perturbative expansion
in $\g$. However, one must be careful, because this fails at small $r$, since, as follows from (23), in this regime $\sqrt\g$ appears at
the denominator.

The RN metric is known to exhibit a singularity at the origin and two horizons at
$$\tilde r_\pm=M\pm\sqrt{M^2-P^2-Q^2}.\eqno(25)$$
Unfortunately, it is not possible to obtain an exact expression for the location of the horizons of the metric (21).
We can obtain an approximation by expanding in the small parameter $\g$,
$$e^{2\n}=1-{2M\over r}+{P^2+Q^2\over r^2}-{\g^2Q^2\over5r^6}+o(\g^4).\eqno(26)$$
The leading-order corrections are proportional to $\g^2$, and we can
compute the zeroes of the metric as $r_\pm=\bar r_\pm+\g^2\D r_\pm+o(\g)$, where
$$\D r_\pm=\pm{\,Q^2\over5\,\br_\pm^4(\br_+-\br_-)}.\eqno(27)$$
Hence, the two horizons are farther than in the RN case.
From (27) follows that at leading order the condition of extremality $r_+=r_-$ is, recalling that $\g^2=3\ba P^2$,
$$M^2\approx Q^2+P^2+{3\ba P^2Q^2\over5(P^2+Q^2)^2}.\eqno(28)$$
The causal structure is analogous to that of RN: for $M$ greater than its extremal value, one has two horizons, while for $M$ smaller than the
extremal value a naked singularity occurs. Although these results are obtained for small $\g$, the qualitative behavior of the solution remains
the same also for generic values of the coupling constant.

The thermodynamical quantities can be computed in the standard way:
the area of the external horizon is usually identified with the entropy, therefore
$$S=4\p\(\br_+^2+{6\ba P^2Q^2\over5r_+^3(\br_+-\br_-)}\)+o(\ba^2),\eqno(29)$$
while the temperature can be calculated as
$$T={1\over4\p}{e^{2\n}\over dr}\Big|_{r=r+}={1\over4\p}\({\br_+-\br_-\over\br_+^2}+{6\ba P^2Q^2\over5\br_+^7}\,{2\br_+-\br_-\over \br_+-\br_-}\)+o(\ba^2).\eqno(30)$$
For extremal black holes the temperature vanishes.
Both temperature and entropy are increased \wrt the \RN black hole.

Let us briefly comment on the case $\a<0$. The metric function has a simpler form,
$$e^{2\n}=\ 1-{2M\over r}+{P^2\over r^2}+{Q^2\over2\sqrt\g\,r}\[{\p\over2}-\arctan{r\over\sqrt\g}-\ha\log{r-\sqrt\g\over r+\sqrt\g}\],\eqno(31)$$
where now $\g=\sqrt{3|\ba|P^2}$, but has the same \ab (22) as the previous solution. Also the expansion for small $\g$ has the same form as (26)
except for the sign of the term proportional to $\g^2$. The curvature scalar is $R={4\g^2Q^2\over(r^4-\g^2)^2}$.
Now a curvature singularity occurs at the surface $r_0=\sqrt\g$, while the horizons are located at
$$r_\pm\approx\br_\pm\pm{3\ba P^2Q^2\over5\,\br_\pm(\br_+-\br_-)}.\eqno(32)$$
If $r_0>r_-$, a single horizon is present
and the causal structure is similar to that of the \sch solutions. Otherwise, the properties are analogous to those of the solution with positive $\a$
and all the previous formulas still hold, taking into account that $\ba$ has opposite sign. This is true in particular for the thermodynamical quantities.
\goodbreak

\section{4. Conclusion}
We have considered the effect of the nonlinearity of the electrodynamics induced by a five-dimensional KK model with
Einstein-GB lagrangian on the dyonic solutions with a pointlike source. While it is well known that purely electric or magnetic solutions
are not modified in this model, we have shown that the dyonic solutions differ from those of the Maxwell theory, and the electric field can be
regular everywhere.

In our model, the \fe contain at most cubic terms in $A_i$, but the model can be generalized to higher powers by increasing the
number of the internal dimensions and adding higher-order GB terms.
Also in this case the pure electric or magnetic fields of pointlike sources maintain the standard form, but the
dyonic solutions are modified and for suitable ranges of values of the coupling constants the singularity of the electric field is suppressed.

We have also examined the coupling with gravity and have found a new class of solutions that modify the RN metric, with a Maxwell field identical
to the flat space solution and a metric that deforms the RN solution. The solutions still depend on the three parameters $M$, $Q$ and $P$, but
are no longer dual for the interchange of $Q$ and $P$.
For positive $\a$ they exhibit a pointlike singularity, while for negative $\a$ the singularity is spherical.
The horizon structure is similar to that of RN, with two horizon, but for some values of the parameters it can present one or no horizons.

Our result is notable since it shows that the introduction of nonlinear equations for the electromagnetic fields affects the results of the \bh
uniqueness theorems also in case of minimal coupling to gravity, analogously to what happens in more general models [13-15].

Going to higher dimensions also allows the introduction of Yang-Mills fields through the Kaluza-Klein mechanism. Of course, in this case
more complicated solutions are expected.

In this paper we have not considered the nonminimal coupling between gravity and electromagnetism induced by the KK-GB model.
Such coupling spoils the possibility of solving the \fe analytically, but the main properties of the solutions should be preserved. We plan to investigate this topic in a future publication [22].

\bigbreak
\beginref
\ref [1] W. Heisenberg and H. Euler, Z. Phys. {\bf 98}, 714 (1936)    .
\ref [2] M. Born and L. lnfeld, \PRS{A144}, 435 (1934).
\ref [3] J. Pleba\'nski, {\it Lectures on non-linear electrodynamics}, NORDITA, Copenhagen, 1968.
\ref [4] G.W. Horndeski, \JMP{17}, 1980 (1976).
\ref [5] H.A. Buchdal, \JoP{A12}, 1037 (1979).
\ref [6] R. Kerner, C.\ R.\ Acad.\ Sc.\ Paris {\bf 304}, 621 (1987).
\ref [7] F. M\"uller-Hoissen, \PL{B201}, 325 (1998).
\ref [8] H.H. Soleng and \O. Gr\o n, \AoP{240}, 432 (1995).
\ref [9] T. Kaluza, Sitz. Preuss. Akad. Wiss., Math. Phys. {\bf 1}, 966 (1921).
\ref [10] O. Klein, Z. Phys.\ {\bf 37}, 895 (1926).
\ref [11] D. Lovelock, \JMP{12}, 498 (1971).
\ref [12] S.I. Kruglov, \PR{D75}, 117301 (2007).
\ref [13] R. Pellicer and R. J. Torrence, \JMP{19}, 1718 (1969).
\ref [14] H.P. de Oliveira, \CQG{11}, 1469 (1994).
\ref [15] A. Garcia, E. Hackmann, C. L\"ammerzahl and A. Mac\'{\i}as, \PR{D86}, 024037 (2012).
\ref [16] W. Israel, \CMP{8}, 245 (1968).
\ref [17] M. Demia\'nski, Found.\ Phys.\ {\bf 16}, 187 (1986).
\ref [18] G. Gibbons and K. Maeda, \NP{B298}, 741 (1988).
\ref [19] S. Mignemi and N.R. Stewart, \PR{D47}, 5259 (1993).
\ref [20] M.S. Volkov and D.V. Gal'tsov,  JETP Lett.\ {\bf 50} 346 (1989).
\ref [21] J. Schwinger, Science {\bf 165}, 757 (1969).
\ref [22] S. Mignemi, in preparation.
\ref [23] S.I. Kruglov, Grav.\ and Cosm.\ {\bf 27}, 78 (2021).

\end